# Manipulating spin and charge in magnetic semiconductors using superconducting vortices

Mona Berciu[1], Tatiana G. Rappoport[2,3] and Boldizsár Jankó[2,3]

[1]Department of Physics and Astronomy, University of British Columbia, Vancouver, BC V6T 1Z1, Canada

[2] Department of Physics, University of Notre Dame, Notre Dame, IN 46556 USA

[3] Materials Science Division, Argonne National Laboratory, Argonne, Illinois 60439 USA

**The continuous need for miniaturization and increase in device speed[1] exerts pressure on the electronics industry to explore new avenues of information processing. One possibility is to use the spin to store, manipulate and carry information. Indeed, *spintronics* may hold the promise of providing such a new paradigm[2]. However, all spintronics applications are faced with formidable challenges in attempting to find fast and efficient ways to create, transport, detect, control and manipulate spin textures and currents. Here we show how most of these operations can be performed in a relatively simple manner in a hybrid system consisting of a superconducting (SC) film and a paramagnetic diluted magnetic semiconductor (DMS) quantum well (QW). Our proposal is based on the observation that the inhomogeneous magnetic fields of the SC create local spin and charge textures in the DMS, leading to effects such as Bloch oscillations, an unusual Quantum Hall Effect, etc. We exploit the recent progress in manipulating magnetic flux bundles (vortices) in superconductors[3,4] and show how these can create, manipulate and control the spin textures in DMS.**

The magnetic properties of SC and DMS are vastly different. When placed in a magnetic field, a type-II SC allows the field to penetrate in a non-uniform fashion, as Abrikosov vortices – the field lines are bundled up into flux quanta surrounded by superfluid eddy currents[5]. The length scale characterizing the field inhomogeneity is the decay length of the superfluid eddy currents surrounding the core of an isolated vortex. This is known as the penetration depth λ, and is a few tens of nanometers for several SC (*e.g.* λ=39 nm in Nb). In contrast, magnetic impurities (typically Mn) embedded in a *paramagnetic* DMS (such as $Ga_{1-x}Mn_xAs$, *x ~ 1.5 –5%*) align their spins along the external field lines, producing a local magnetization $\mathbf{M(r)} \propto \mathbf{B(r)}$. The strong exchange between these impurities' spins with the spins of free charge carriers present in the DMS induces a local spin polarization of the charge carriers, $\int d\mathbf{r} J(\mathbf{r})\mathbf{M(r)} \cdot \boldsymbol{\sigma} \propto \mathbf{B(r)} \cdot \boldsymbol{\sigma}$. This is incorporated into a formally unchanged Zeeman energy $E_Z = -\frac{1}{2} g_{eff} \mu_B \boldsymbol{\sigma} \cdot \mathbf{B(r)}$, but now with an enormous effective gyromagnetic ratio $g_{eff} \sim 10^2$. The end result is a giant Zeeman splitting between charge carrier spin states aligned parallel and antiparallel to the magnetic field[6].

A novel situation arises when these two materials are brought into close proximity (see Fig. 1a). The inhomogeneous magnetic field of a SC vortex creates a Zeeman potential landscape in the DMS QW. If a large enough field variation occurs on small length scales, magnetic field induced localization of charge carriers occurs in the DMS[7]. These localized states exhibit spin textures, which reflect the local magnetic field distribution. The charge and spin texture remains attached to and adiabatically follows a moving vortex, as long as the Doppler shift in the bound state energy is smaller than the binding energy. Thus, the vortex acts as a *spin and charge tweezer*: control of the vortices' locations and dynamics translates into controlled manipulation of the spin and charge textures in the DMS. A further advantage is that these effects are present in the *paramagnetic* DMS, and therefore a low Curie temperature is not a hindering factor. The devices, however, must operate below the superconducting transition temperature.

To illustrate these concepts, we first consider an isolated vortex in a SC-DMS bilayer structure. This limit is relevant for low fields when vortices are far apart, or when the DMS is covered by nanoscale SC dots with lateral size *d~λ*. The magnetic field[8] of a vortex out of a half-space isotropic SC is used to compute numerically the energies and wave functions of the bound states localized by the vortex field in the DMS QW. For simplicity we use DMS/SC parameters typical of $Ga_{1-x}Mn_xAs$/Nb. We envision the QW as being made from several DMS digital layers[9], which are less disordered than bulk-doped DMS and have the ideal geometry. Good mobilities have been demonstrated in similar QW[10]. Moreover, the charge carrier density can be controlled independently in digital layers. Nb is a canonical type-II SC. It can be grown via sputtering at room-temperature - this limits Mn diffusion, preserving the quality of the QW and preventing magnetic doping of the SC. (For further details on material issues and the numerical solution, see the supplementary material). The results for the local density of states (LDOS), or charge texture, are shown in Fig. 1. For any energy $E_f$ the LDOS is $\left|\Psi_\sigma(r,E_f)\right|^2 = \sum_n |\psi_{n,\sigma}|^2 (r,E)\delta(E_n - E_f)$. (We make the usual assumption that the local tunneling amplitude is a weak function of the energy). The charge distribution bound by the vortex is made predominantly of spin-up contributions, as seen from comparing panels (b) and (c) in Fig. 1: while both spin components extend over the same area (roughly the size of the vortex), their maxima differ by about a factor of 10. The spin-down components appear due to the finite radial component of the vortex field (see inset (d) of Fig. 1). The spin-up components are concentrated right under the vortex core where the perpendicular field is largest, whereas the spin-down components are pushed away from the vortex core, close to where the radial magnetic field is largest. The total charge and spin distributions plotted in Fig. 1 (e) show that the spin texture is strongly polarized along the axis of the vortex. These textures should be observable with spin-polarized scanning tunneling spectroscopy. However, as discussed below, magneto-transport may provide the most direct signature of the appearance of these textures in the DMS.

The location of the vortices can be constrained spatially using nano-engineered grooves in the SC layer: the vortices are trapped in the regions where the SC film is thinnest. The vortex position in the groove can be controlled by sending a current through the SC layer. The current exerts a transversal force which moves the vortex[3], together with the texture it created in the DMS. The spin textures attached to single vortices can therefore be used as operational units, or as building blocks in systems with multiple vortices. A simple device controlling the distance between two spin textures is sketched in Fig. 2 (a), (b). Other examples, ranging from using ratchets to create vortex (and thus spin) lenses and pumps up to quantum cellular automata, are described in the supplementary material.

In uniform SC films there is a simpler way to control the location of the vortices. The repulsion between vortices leads to appearance of regular vortex lattices. Since each vortex carries a magnetic flux $\phi_0/2=h/(2e)$, the distance between vortices can be *continuously tuned* with the external magnetic field. Let us first analyze the quasi one-dimensional (1-D) case, with long stripes of superconducting film, of width $w\sim\lambda$, deposited above a DMS QW. Such SC structures allow formation of 1-D vortex chains. We use the 1-D case to give an intuitive illustration of the various phenomena expected to occur in a tunable SC-DMS hybrid system, but present numerical results only for the more complex 2-D case. The low field limit (Fig. 2(c)) has the spin textures of a 1-D set of isolated vortices. If the external magnetic field is increased, the inter-vortex distance $a(B)$ is reduced, together with the modulation of the Zeeman potential which traps the charges under individual vortices. Consequently, the bound states under each vortex are broadened into bands (Fig. 2(d)), and the trapped charges undergo a delocalization transition: they are now free to propagate along the 1-D array. The new eigenstates of the charge carriers in the DMS are Bloch states of a 1-D crystal, reflecting its translational symmetry. If the vortex array is along the **x** axis, then $\psi_k(x,y,z) = e^{ikx}u_k(x,y,z)$ with $u_k(x,y,z) = u_k(x+a(B),y,z)$ (see the supplementary material for details). An immediate consequence of – or, conversely, the most convincing evidence for – the translational symmetry and the establishment of Bloch eigenstates is the appearance of the so-called *Bloch oscillations*: If the DMS is subjected to an electric field $\mathbf{E} = E\hat{\mathbf{x}}$, then *A/C* charge oscillations should appear, with a characteristic period[11] $T=h/[eEa(B)]$. Bloch oscillations have never been observed in normal metals, due to scattering off impurities before the Bloch states traverse the entire k-space Brillouin zone of size $2\pi/a(B)$. They have been detected only in artificial heterostructures where the lattice constant $a$ is engineered to be much larger than in crystals. The structure we suggest is unique because in this case the periodicity $a(B)\sim B^{-1}$ is tunable with the external magnetic field, and therefore Bloch oscillations can be searched for more easily.

Even more interesting is the 2-D case. For an external magnetic field $B>B_{c1}$, the SC vortices order on a triangular lattice[5]. Since each unit cell encloses the magnetic flux $\phi = Ba^2\sqrt{3}/2 = \phi_0/2$ of a vortex, the lattice constant $a \sim 1/\sqrt{B}$ is controlled by $B$. The 2D electron gas in the DMS QW is now in a periodic Zeeman potential and a perpendicular magnetic field. Its energy spectrum has a fractal-like structure as a function of $\phi/\phi_0$ (named the Hofstadter butterfly because it resembles a butterfly). Since $\phi/\phi_0=q/p=1/2$, the energy bands are those of a $q=1$, $p=2$ triangular Hofstadter butterfly[12,13]. In the limit $a(B)>>\lambda$, the vortices are well-separated. The "atom-like" states bound to each isolated vortex widen into energy bands, typical of an ordered crystal. The width of the bands is defined by the hopping amplitude $t$ between neighboring bound states. If $t$ is small compared to the energy spacing between consecutive bound states, this Hofstadter problem corresponds to the regime of a dominant periodic modulation, which can be treated within a simple tight-binding model[12] and each band splits into $p$ (here $p=2$) magnetic subbands. This band structure has unique signatures in the magnetotransport, as discussed below. Its measurement would provide a clear signature of the Hofstadter butterfly, which is a matter of considerable experimental interest[14-16]. As the external field $B$ is increased such that $a\sim\lambda$, the magnetic fields of different vortices start to overlap significantly. In this limit, the total magnetic field in the DMS layer is almost homogeneous. Its average is $B\hat{z}$ and it has a small additional periodic modulation. For such quasi-homogeneous magnetic fields, the Zeeman interaction can no longer induce trapping; instead it reverts to its traditional role of lifting the spin degeneracy. The system still corresponds to a $\phi/\phi_0=1/2$ Hofstadter butterfly, but now in the other asymptotic limit, *i.e.* that of a weak periodic modulation. In this case, each Landau level splits into $q$ subbands, with a bandwidth controlled by the amplitude of the weak modulation. The case $\phi/\phi_0=1/2$ has $q=1$, *i.e.* there are no additional gaps in the band structure. One therefore expects to see the usual Integer Quantum Hall Effect (IQHE) in magnetotransport in this regime[17].

 Three aspects must be emphasized: (i) to our knowledge, this is the first proposal of an experimental setup for the measurement of the Hofstadter butterfly in the limit of a *strong* periodic modulation (the tight-binding limit); (ii) the applied field $B$ plays a very unusual role here. Typically, one assumes a fixed lattice cell. Varying $B$ changes the flux $\phi$ through the unit cell, resulting in the self-similar eigenspectrum of the Hofstadter butterfly. In contrast, here a change in $B$ implies a change in the lattice constant, but $\phi$ *always* equals $\phi_0/2$. As long as there is a vortex lattice, the setup corresponds to a $\phi/\phi_0=1/2$ Hofstadter butterfly irrespective of the value of *B*. Instead, *B* controls the amplitude of the periodic modulation, from being the large energy scale (small *B*) to being a small perturbation (large *B*); (iii) here the spin plays an essential role, since the periodic modulation is mainly due to the spin-dependent Zeeman potential which is strongly enhanced in the DMS. In previous work considering periodically modulated magnetic fields[18-20] the Zeeman term was ignored. The spin-polarized bound states we discuss cannot appear in the absence of the Zeeman term, and the physics is qualitatively different.

We now present numerical results that confirm these expectations (for details, see the supplementary material). In Fig. 3 we show the evolution of the charge carrier energy spectra as *B* is varied. The large-*B* limit corresponds to a small periodic modulation and indeed we see the emergence of equally spaced Landau levels with a weak dispersion due to the weak modulation. As *B* is lowered, the band structure evolves into that corresponding to a strong modulation, and for the lowest *B* we see flat bands corresponding to isolated vortex bound levels. This significant change in the dispersion as *B* is varied is reflected in magnetotransport, in particular in the Hall conductivity. When Fermi level is inside a gap, the Hall conductivity has a plateau at a quantized value[21] $\sigma_{xy}=ne^2/h$. Laughlin's gauge arguments[22] insure that this quantization holds in the presence of moderate disorder (small enough to not close these gaps). In the weak-modulation limit $\sigma_{xy}$ increases by $e^2/h$ after each transition through a Landau level, as in typical IQHE. In the tight-binding limit, it is known that the Hall conductance of the entire band is always zero, although if inner gaps are opened by the periodic modulation, they can have different Hall conductances. By explicitly counting the number of zeros (and their vorticities) of the Bloch wave function $u_{\mathbf{k}}(\mathbf{r}) = e^{-i\mathbf{k}\cdot\mathbf{r}}\psi_{\mathbf{k}}(\mathbf{r})$ inside the magnetic Brillouin zone[23-25], we have verified that the Hall conductance of the sub-bands follow this expected behavior in both limits of the Hofstadter problem (see Fig. 3).

Fig. 4 shows $\sigma_{xy}$ as a function of the charge carrier density n and magnetic field *B*. Disorder leads to some widening (arbitrarily drawn in this figure) of each gap, through localization of some fraction of states. The main sources of disorder are inhomogeneity in the DMS and vortex lattice irregularities. The former can be minimized using digital doping, leading to reasonably high mobilities[10]. The latter and the so-called Bean profile can be practically eliminated by using SC with low intrinsic pinning. As *B* is varied at constant n (or vice-versa) one expects to see non-trivial sequences of the plateau values of $\sigma_{xy}$. Experimental measurement of such nontrivial sequences would clearly signal the emergence of the tight-binding limit of the Hofstadter butterfly, correlated with the appearance of the spin and charge textures in the DMS QW.

In conclusion, we demonstrated that SC vortices can act as effective spin and charge tweezers in a DMS QW. Recent progress in manipulation of SC vortices can be directly used to create spin lenses, ratchets, pumps, cellular automata, and other devices. The spin textures can also be placed on regular lattices that can be tuned by modifying the external field. This provides a new system to investigate many beautiful *basic physics* phenomena, such as Bloch oscillations or the $\phi/\phi_0=1/2$ Hofstadter butterfly, but also 1-D and 2-D metal-insulator transitions, to name just a few. Thus, hybrid systems of DMS and SC have a remarkable potential both in applied and basic condensed matter physics.

**Supplementary information** accompanies the paper on Nature's website (http://www.nature.com)

Correspondence and requests for material should be addressed to B. Jankó (bjanko@nd.edu)

**Acknowledgments:** We thank T. Wojtowicz and A. A. Abrikosov for useful discussions. This research was supported by NSERC and the Research Corporation (M.B.), by NSF NIRT (T.G.R and B.J.) and by the Alfred P. Sloan Foundation (B.J.). We are grateful for the hospitality of the Argonne National Laboratory where parts of this work were carried out.

**Competing interests statement.** The authors declare that they have no competing financial interests.

**References**
1. Peercy, P. S. The drive to miniaturization. *Nature* **406**, 1023–1026 (2000).
2. Wolf, S. A. *et al.* Spintronics: A spin-based electronics vision for the future. *Science* **294**, 1488–1495 (2001).
3. Lee, C. S., Janko, B., Derenyi, I. & Barabasi, A. L. Reducing vortex density in superconductors using the 'ratchet effect'. *Nature* **400**, 337–340 (1999).
4. Savel'ev, S. & Nori, F. Experimentally realizable devices for controlling the motion

of magnetic flux quanta in anisotropic superconductors. *Nature Materials* **1**, 179–184 (2002).
5. Abrikosov, A. A. On the Magnetic Properties of Superconductors of the Second Group. *Soviet Physics JETP* **5**, 1174-1182 (1957).
6. J. K. Furdyna, Diluted magnetic semiconductors *J. Appl. Phys.* **64**, R29-R64 (1988).
7. Berciu, M. & Janko, B. Nanoscale Zeeman localization of charge carriers in diluted magnetic semiconductor-permalloy hybrids. *Physical Review Letters* **90**, 246804 (2003).
8. Pearl, L. Current distribution in superconducting films carrying quantized fluxoids. *Applied Physics Letters* **5**, 65-66 (1964).
9. Kawakami, R. K. *et al.* (Ga,Mn)As as a digital ferromagnetic heterostructure. *Applied Physics Letters* **77**, 2379-2381 (2000).
10. Jaroszynski, J. *et al.*, Ising Quantum Hall Ferromagnet in Magnetically Doped Quantum Well, *Physical Review Letters* **89**, 266802 (2002).
11. Bloch, F. The quantum mechanics of electrons in crystal lattices *Z. Phys.* **52**, 555-600 (1928).
12. Hofstadter, D. R. Energy-levels and wave-functions of bloch electrons in rational and irrational magnetic-fields. *Physical Review B* **14**, 2239–2249 (1976).
13. Claro, F. H. & Wannier, G. H. Magnetic subband structure of electrons in hexagonal lattices. *Physical Review B* **19**, 6068–6074 (1979).
14. Melinte, S. *et al.* Laterally modulated 2D electron system in the extreme quantum limit. *Physical Review Letters* **92**, 036802 (2004).
15. Schlosser, T. *et al.* Internal structure of a Landau band induced by a lateral superlattice: A glimpse of Hofstadter's butterfly. *Europhysics Letters* **33**, 683–688 (1996).
16. Albrecht, C. *et al.* Evidence of Hofstadter's fractal energy spectrum in the quantized Hall conductance. *Physical Review Letters* **86**, 147–150 (2001).
17. von Klitzing, K., Dorda, G. & Pepper, M. New Method for High-Accuracy Determination of the Fine-Structure Constant Based on Quantized Hall Resistance. *Physical Review Letters* **45**, 494-497 (1980).
18. Bending, S. J., von Klitzing, K. & Ploog, K. Weak localization in a distribution of magnetic-flux tubes. *Physical Review Letters* **65**, 1060–1063 (1990).
19. Nielsen, M. & Hedegard, P. Two-dimensional electron transport in the presence of magnetic flux vortices. *Physical Review B* **51**, 7679-7699 (1994).
20. Chang, M. C. & Niu, Q. Electron band structure in a two-dimensional periodic magnetic field. *Physical Review B* **50**, 10843-10850 (1994).
21. Thouless, D. J., Kohmoto, M., Nightingale, M. P. & Dennijs, M. Quantized hall conductance in a two-dimensional periodic potential. *Physical Review Letters* **49**, 405–408 (1982).
22. Laughlin, R. B. Quantized hall conductivity in 2 dimensions. *Physical Review B* **23**, 5632–5633 (1981).
23. Kohmoto, M. Topological invariant and the quantization of the hall conductance. *Annals Of Physics* **160**, 343–354 (1985).
24. Usov, N. A. Theory of the quantum hall effect in a two-dimensional periodic potential. *Soviet Physics JETP* **67**, 2565-2573 (1988).

25. Demikhovskii, V. Y. & Khomitskiy, D. V. Quantum Hall effect in a p-type heterojunction with a lateral surface superlattice. *Physical Review B* **68**, 165301 (2003).

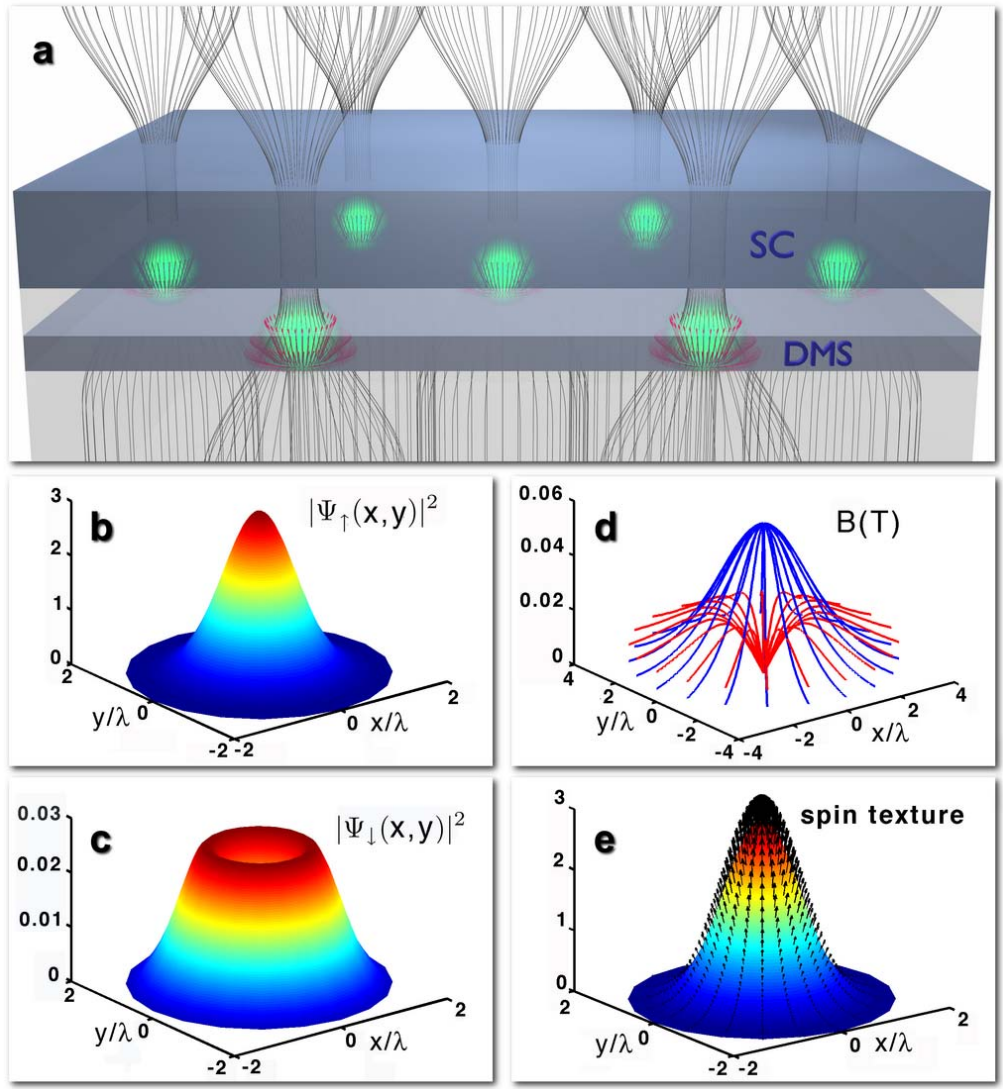

**Figure 1: Spin and charge textures created by SC vortices in a DMS.** (a) Sketch of the hybrid structure, containing a SC layer in the vortex phase and a DMS QW. The field inhomogeneity of the vortex state is imprinted onto the DMS layer. Spin and charge textures are trapped in the high-field regions inside the DMS; (b) charge texture for spin-up states; (c) charge texture for spin-down states (note the smaller scale); (d) spatial variation of the axial (blue) and radial (red) components of the isolated vortex field; (e) total spin-charge texture. Arrows indicate the direction and relative magnitude of the local spin in the texture. (b)-(e): The DMS-SC distance is $z$=10 nm and we use typical values of $\lambda \sim$ 40 nm and $\xi \sim$ 35 nm for the SC, and $m$=0.5$m_e$ and a moderate $g_{eff}$=500 for the DMS charge carriers.

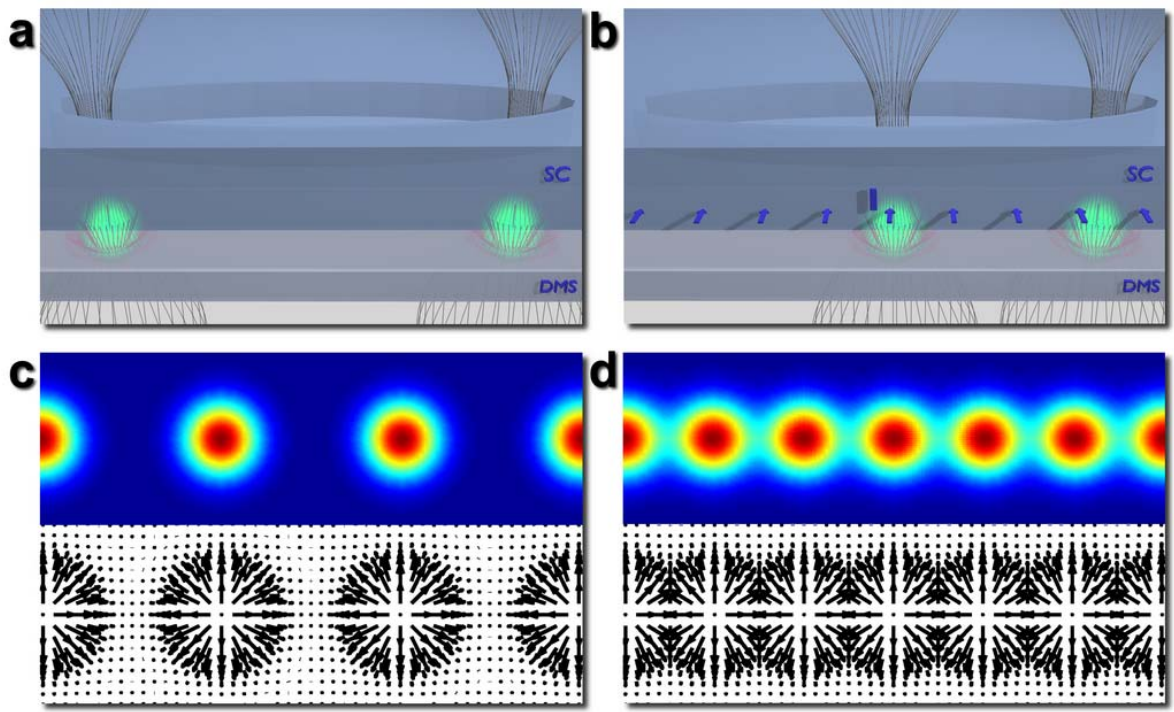

**Figure 2: Devices with tunable distance between spin-charge textures** (a) A finite-size groove is produced in the SC film by etching and two vortices are trapped into it. Due to repulsion, in equilibrium the vortices will stay at maximum possible separation; (b) if a current flows through the SC layer, its transversal force on the vortices will move them (and their spin textures) closer together; (c) Field (top panel) and spin texture (bottom panel) for a one-dimensional vortex array in low field regime, when inter-vortex separation $a(B)$ is larger than the penetration depth, $a(B) > \lambda$. Vortices trap bound states and form textures independently of each other; there is little overlap between the resulting spin and charge textures; (d) Same as in (c), but now for intermediate-to-high magnetic field regime: there is substantial overlap between the bound states, resulting in a collective spin and charge texture.

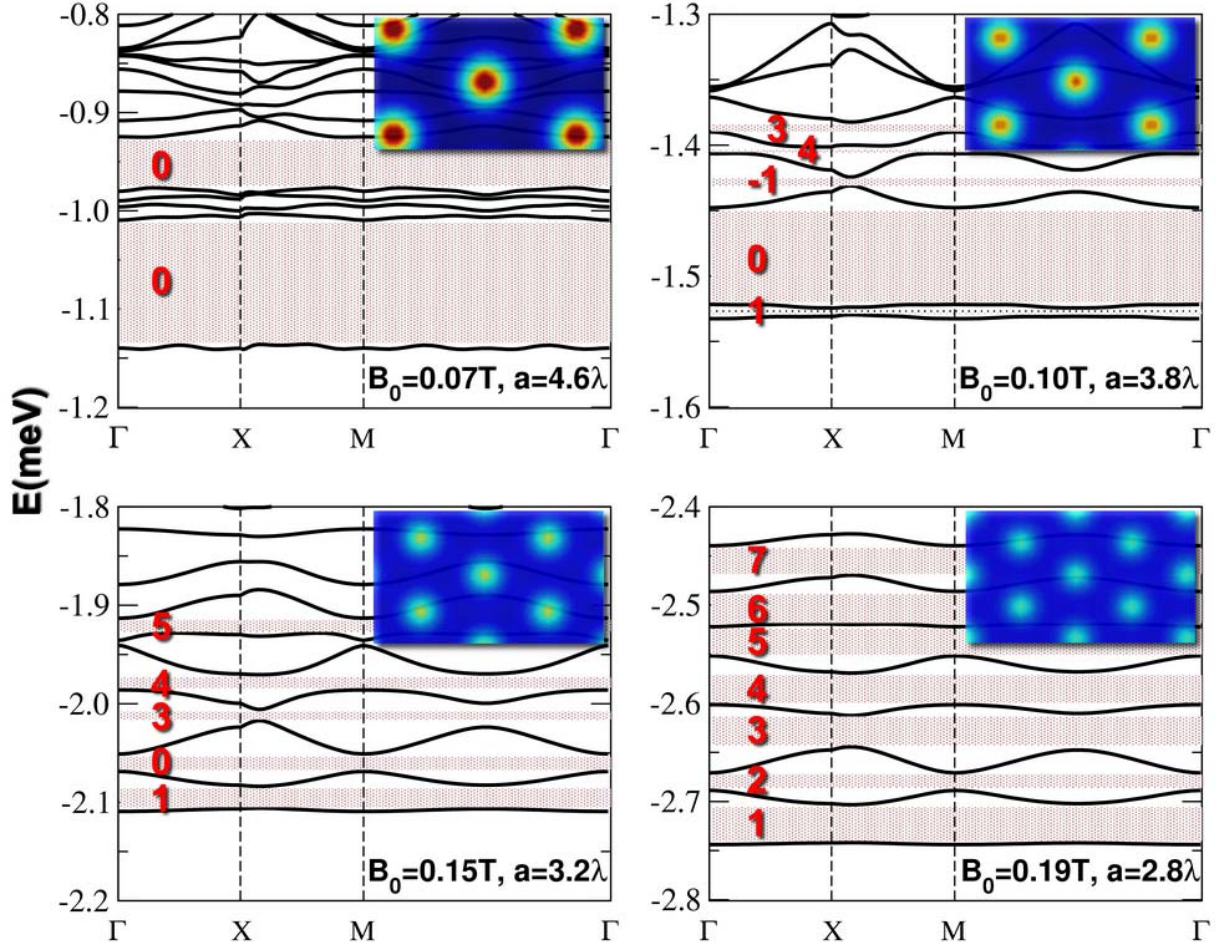

**Figure 3: Band structure of the DMS in the hybrid system for $B$=0.07, 0.10, 0.15 and 0.19 T**. The symmetry points in the Brillouin zone are defined as usual: $\Gamma=(0,0)$; $M=(\pi/a,0)$ and $X=(\pi/a,\pi/a)$. The shaded regions mark the gaps. The value $n$ for the expected plateau in the Hall conductivity, $\sigma_{xy}=ne^2/h$, is marked for the first few gaps. The parameters are as in Fig. 1, except for $z$=8 nm. The insets show the modulation of the magnetic field in the DMS. The number of vortices per area increases for increasing magnetic fields.

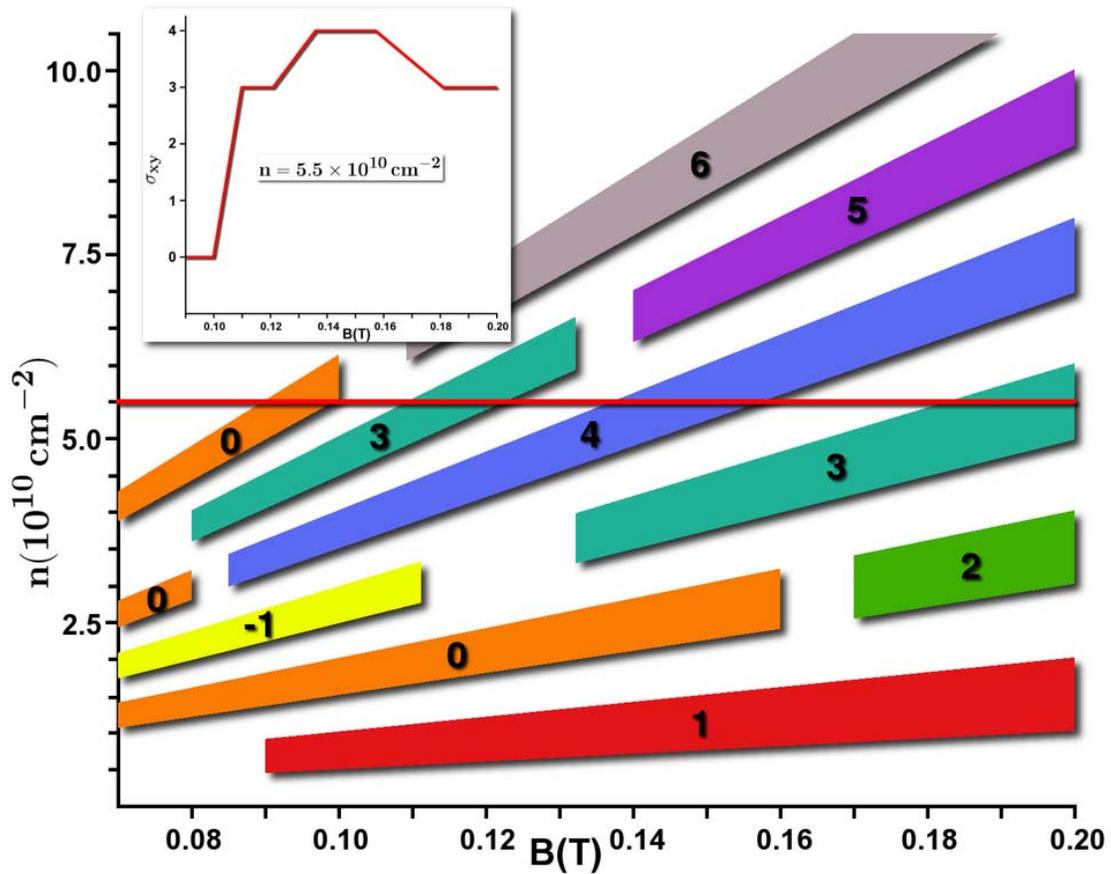

**Figure 4: Hall conductivity $\sigma_{xy}$ in units of $e^2/h$, as a function of the magnetic field $B$ and the charge carrier density n**. The colored areas mark the gaps between the bands. For some ranges of $B$, neighboring bands touch and some of the gaps close. The integers give the quantized values of the $\sigma_{xy}$ plateaus. For example, $\sigma_{xy}$ as a function of $B$, at a constant density n=5.5×10$^{10}$cm$^{-2}$(red line) is shown in the inset. It is quantized every time the Fermi level is inside a gap. The parameters are the same as for Fig. 3.